\documentclass[%
 reprint,
 amsmath,amssymb,
 aps,
]{revtex4-1}
 
\usepackage{graphicx}
\usepackage{dcolumn}
\usepackage{bm}
\usepackage{comment}
\usepackage{color}

\begin{document}

\preprint{APS/123-QED}

\title{Molecular shear heating and vortex dynamics in thermostatted two-dimensional
Yukawa liquids}
\author{Akanksha Gupta}
\altaffiliation{Institute for Plasma Research, India}

 \author{Rajaraman Ganesh}%
 \email{ganesh@ipr.res.in}

\affiliation{%
 Institute for Plasma Research, HBNI, Bhat, Gandhinagar - 382428, India
 }%
\author{Ashwin Joy}
\affiliation{
 Department of Physics, Indian Institute of Technology Madras, Chennai - 600036, India
}%

\begin{abstract}
It is well known that two-dimensional macroscale shear flows are susceptible to instabilities leading to macroscale vortical structures. The linear and nonlinear fate of such a macroscale flow in a strongly coupled medium is a fundamental problem. A popular example of a strongly coupled medium is a dusty plasma, often modelled as a Yukawa liquid. Recently, laboratory experiments and MD studies of shear flows in strongly coupled Yukawa liquids, indicated occurrence of strong molecular shear heating, which is found to reduce the coupling strength exponentially leading to destruction of macroscale vorticity. To understand the vortex dynamics  of strongly coupled molecular fluids undergoing macroscale shear flows and molecular shear heating,  MD simulation has been performed, which allows the macroscopic vortex dynamics to evolve while at the same time, ``removes" the microscopically generated heat without using the velocity degrees of freedom. We  demonstrate that by using a configurational thermostat in a novel way, the microscale heat generated by shear flow can be thermostatted out efficiently without compromising the large scale vortex dynamics.  In present work, using MD simulations, a comparative study of shear flow evolution in Yukawa liquids in presence and absence of molecular or microscopic heating is presented for a prototype shear flow namely, Kolmogorov flow.
\end{abstract}

\pacs{Valid PACS appear here}
\maketitle


\section{\label{sec:level1}Introduction}

\noindent Micron-sized dust grains get highly charged when immersed in a conventional  plasma because of continuous bombardment of electrons \citep{rao}. The average charge on a single dust grain is typically  $\sim10^{3}e-10^{4}e$, where $e$ is the absolute electronic charge.  There are many examples of plasma  in nature wherein, large sized grains interact with an ambient plasma and play an important role. For example, comets, planetary rings, white dwarf, earth's atmosphere and in laboratory conditions such as plasma processing reactors, plasma torch and fusion devices \citep{fusion}. \\

\noindent These charged grains or dust particles interact via a shielded Coulomb interaction or Yukawa potential as the ambient plasma shields the grain charge. Such plasmas can be characterized by two  non-dimensional parameters $\kappa=a/\lambda_{D}$ (where $a$ is inter-grain-spacing and $\lambda_{D}=\lambda_{i}\lambda_{e}/\sqrt{\lambda_{i}^{2}+\lambda_{e}^{2}}$ is  Debye length of background plasma, $\lambda_{i}$, $\lambda_{e}$ are Debye length of electron and ion respectively) and coupling parameter $\Gamma = Q_{d}^{2}/(4\pi\varepsilon_{0}ak_{B}T_{d} )$ wherein $Q_{d}$  and $T_{d}$ are charge and temperature  of grain. High charge on grain enhances the grain-grain  potential energy and hence the coupling parameter $\Gamma$. Depending upon $\kappa$ and $\Gamma$ values, such strongly coupled grains can exhibit  solid-like (for $\Gamma > \Gamma_m$)$\:$\citep{goree,li,morfill}, liquid-like (for $1 < \Gamma < \Gamma_m$)$\:$\citep{vlad} and gas-like (for $\Gamma << 1$)\citep{rao} features. Here $\Gamma_m$ is the liquid to solid phase transition point\\

\noindent In laboratory  experiments, grain dynamics can be visualized (by  unaided eye)  and tracked  by optical cameras \citep{goree}. Using MD simulation, where the interaction between dust grain is modelled by shielded Coulomb or Yukawa potential, various properties of  grain medium, such as phase transition, instability, transport,  grain  crystallization, determination of transport coefficients such as shear and bulk viscosities \cite{shear_MD}, Maxwell relaxation time \citep{Ashwin_tau_PRL}, heat conduction, wave dispersion \cite{wave_disper}, self diffusion \cite{diffusion_MD}, fluid instability such as shear driven, Kelvin-Helmholtz instability \citep{ashwin}   has been addressed. In recent past, strong peaks of temperature have been observed at the velocity shear location in macroscopic shear flows of grain generated by external laser-drive in Yukawa liquid experiments \citep{laser_PRE,yan} and also  in MD simulations \citep{Aka, ashwin_coevolution}. It was seen that shear flows in Yukawa liquids lead to generation of heat
at the microscopic scales  resulting into the formation of a heat front \cite{ashwin_coevolution}. Using Kolmogorov flow \citep{sinai,obu,landau,kelley,bena}  as a initial shear flow in Yukawa liquids, it was also demonstrated that molecular shear heating destroys the vortex dynamics and reduce the coupling strength exponentially \citep{Aka}. \\

\noindent In the present work, we investigate whether or not it is possible to address macroscale vortex dynamics using MD simulation and  at the same time maintain the grain bed at the desired temperature. In the following, we consider a prototype shear flow namely Kolmogorov flow, which has been studied in the context of laminar to turbulent transition \citep{Aka} wherein it was shown that the average coupling strength decreases exponentially with time due to molecular shear heating. We propose and demonstrate here that using MD simulation and thermostat based on configurational space degree of freedom  \citep{Rugh1997,Butler1998,Braga2005,BT2008} (also called profile unbiased thermostat or PUT), it is possible to ``remove" heat yet study macroscale vortex dynamics. Below we show the correspondence between PUT and a velocity scaling or profile based thermostat (PBT) for the case of Kolmogorov shear flows.\\ 

\noindent In the present work, we have introduced the method of thermostatting namely, configurational thermostat. In the past, Rugh \cite{Rugh1997} and Butler\cite{Butler1998} presented a method of calculating the temperature of a Hamiltonian dynamical system. This method of calculating temperature only depends upon the configurational information of the system, hence named configurational temperature. Influenced by the concept of configurational temperature a new method of thermostatting namely configurational thermostat, has been introduced by  Delhommelle and Evans \citep{Delhommelle2001,Delhommelle2002}, Patra and Bhattacharya \citep{patra2014} and,  Braga and Travis \citep{Braga2005,BT2008}. Due to its relative simplicity in implementation, we have chosen the Braga-Travis version of PUT. This amounts to invoking appropriate Lagrange multipliers that will efficiently couple the grain bed to a configurational thermostat. Using this PUT we demonstrate that the average coupling strength can be controlled without compromising the effects of strong correlations on the macroscopic shear flow and vortex dynamics. A detailed comparison of the evolution and dynamics of Kolmogorov flow in the presence and absence  of molecular shear heating, its effect on linear growth rate, non-linear saturation and transition from laminar to turbulence flow will be presented.\\

\noindent This paper is presented in the following fashion.
In Sec. \ref{CongiT} and \ref{Thermostat}, we describe configurational temperature and configurational thermostat used in our work. In Sec. \ref{heat}, molecular shear heating effects on shear flow in presence and absence of shear heating is described. In the last section Sec.\ref{sec.conclusion}, we conclude  and  discuss about future  work.

\section{Configurational Temperature }
\label{CongiT}
\noindent Thermodynamic temperature is a estimate of the average or random kinetic energy of the particles in a system.  According to kinetic theory of gases, the kinetic temperature can be expressed as:\\
\begin{equation}
k_{B}T_{kinetic}= \frac{1}{Nd}\sum_{i=1}^{N}\frac{m_{i}v_{i}^{2}}{2}
\label{kinetic}
\end{equation}

\noindent where $k_{B}$ is Boltzmann constant and $m_{i}$, $v_{i}$ are mass and instantaneous velocity of $``i^{th}"$ particle respectively, $N$ and $d$ are number of particles and dimensionality of the system. Apart from this definition of temperature, assuming ergodicity, one can also define temperature by a purely dynamical time averaging of a function which is related to the curvature of energy surface. For example, H. H. Rugh \citep{Rugh1997}, in 1997, presented a dynamical approach for measuring the temperature of a Hamiltonian dynamical system. Rugh argues, using statistical thermodynamics for any closed Hamiltonian system, that the kinetic and configurational temperature would be asymptotically identical for a closed system. His definition of configurational temperature is:
\begin{equation}
\frac{1}{k_{B}T_{config}}=\boldsymbol{\nabla}\cdot\dfrac{\boldsymbol{\nabla} H}{\Vert\boldsymbol{\nabla} H\Vert^{2}}
\label{Rugh}
\end{equation}
where $H$ is the Hamiltonian of classical dynamical system. 
Butler \citep{Butler1998} and others \citep{han,Branka2011}  generalised Rugh's idea for any function $\boldsymbol{B}(\boldsymbol{\Lambda})$ of phase space, such that 
\begin{equation}
k_{B}T = \Big \langle \frac{\boldsymbol{\nabla} H(\boldsymbol{\Lambda})\cdot \boldsymbol{B}(\boldsymbol{\Lambda})}{\boldsymbol{\nabla} \cdot\boldsymbol{B}(\boldsymbol{\Lambda})}\Big \rangle 
\label{branka}
\end{equation}
where $\boldsymbol{\Lambda}= \lbrace q_{1},q_{2}......q_{3N}, p_{1},p_{2}......p_{3N}\rbrace$ is the phase space vector and ($q_{j}, p_{j}$) are $6N$ generalized coordinates for  conjugate positions and momenta respectively. The Hamiltonian of the system $H(\boldsymbol{\Lambda})=K(p_{j})+U(q_{j})$, where $K(p_{j})$ and $U(q_{j})$ are kinetic and potential energy of the dynamical system respectively. In Eq.$\:$\ref{branka}, $B(\boldsymbol{\Lambda})$ can be any continuous and differentiable vector in phase space. 
For example, if one chooses  $\boldsymbol{B}(\boldsymbol{\Lambda})$ as $\boldsymbol{B}(\boldsymbol{\Lambda}) = \boldsymbol{B}(p_{1}, p_{2}, ... , p_{3N}, 0, 0, ... , 0)$, Eq.$\:$\ref{branka} yields the familiar kinetic temperature  namely, $(3 N / 2)k_{B}T = \sum_{i=1}^{N}p_{i}^{2}/(2m_{i})$. Similarly when $\boldsymbol{B}(\boldsymbol{\Lambda}) = \boldsymbol{B}(0, 0, ... , 0, q_{1}, q_{2}, ... , q_{3N})=-\boldsymbol{\nabla} U$ (say), Eq.$\:$\ref{branka} gives configurational temperature in potential form 
\begin{equation}
k_{B}T_{config} = \Big \langle \frac{(\boldsymbol{\nabla} U)^{2}}{\nabla^{2} U}\Big \rangle 
\label{main}
\end{equation} 
In a closed  system, these kinetic and configurational temperatures are expected to be asymptotically identical. 
\subsubsection*{Kinetic and configurational temperatures of strongly coupled Yukawa liquid}
\begin{figure}[h!]
\includegraphics[width=3.4in,height=3.2in]{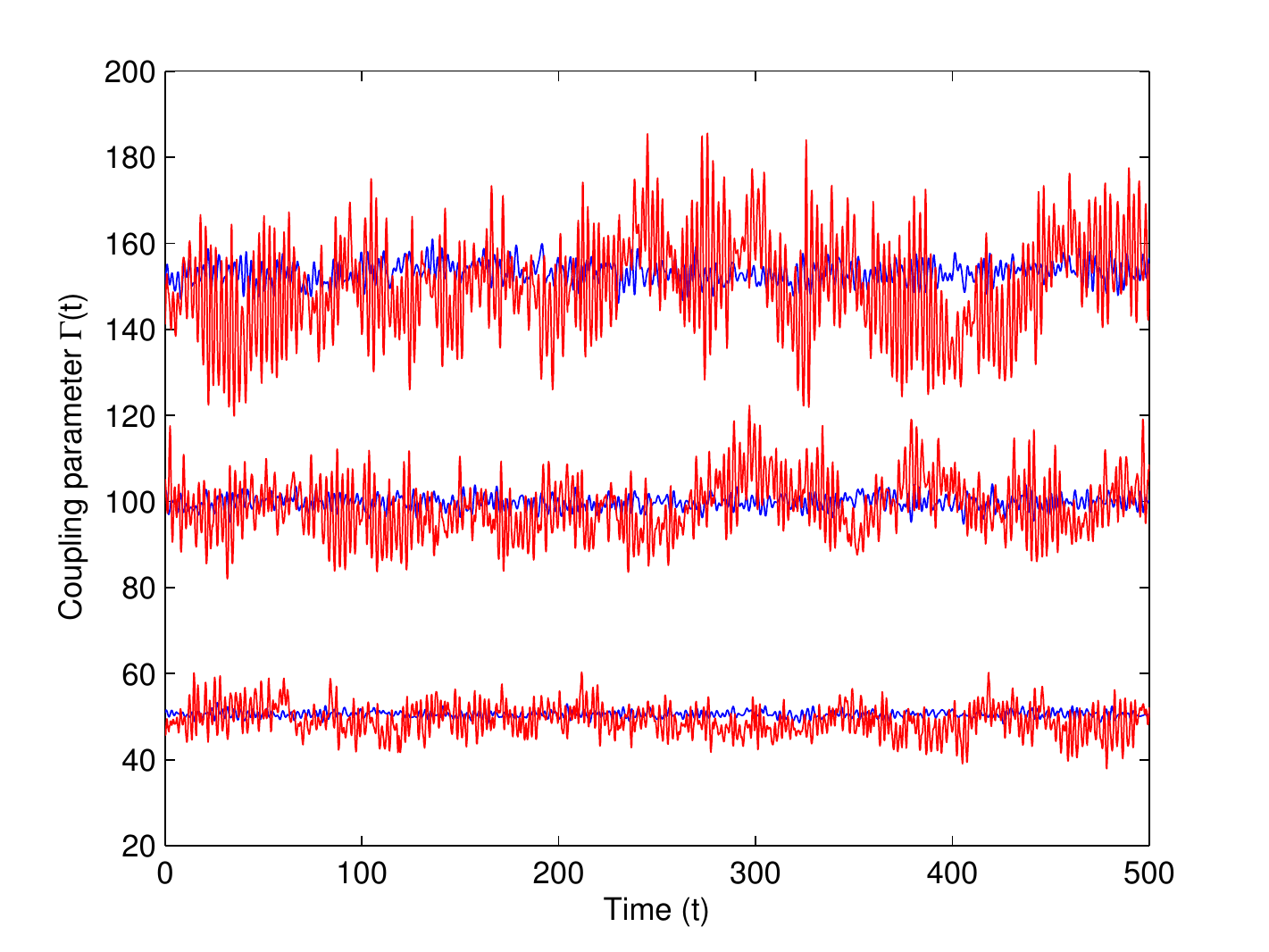}
\caption{color online: Kinetic (blue) and configurational (red) $\Gamma(t)$ extracted under micro-canonical conditions for the Yukawa liquid previously equilibrated at $\Gamma = 50 ,\ 100 ,\ 150$ with $\kappa = 0.5$}
\label{Tc}
\end{figure}

\noindent Due to slow dynamics of grain medium, we consider ambient plasma  properties to be invariant and model only grain dynamics, which is strongly coupled. We first calculate the configurational temperature of the grain medium where, grains interact via screening Coulomb or Yukawa potential :
 \begin{equation}
	  U(\boldsymbol{r}_{i}) = \frac{Q_{d}^2}{4 \pi \epsilon_0 }  \sum \limits_{j\neq i}^N \frac{e^{-r_{ij}/\lambda_D}}{r_{ij}}
	  \label{Yukawa-pot}
	\end{equation}\\
where $r_{ij}=\vert \boldsymbol{r}_{i}-\boldsymbol{r}_{j}\vert$ is the inter particle distance of $i^{th}$ and $j^{th}$ particle. The $N$-body problem is then numerically integrated using our parallelized MD code \cite{AshwinPRE}. We have calculated configurational temperature by using  Eq.$\:$\ref{main} where, interaction potential $U(\boldsymbol{r}_{i})$ is given by Yukawa potential as in Eq.$\:$\ref{Yukawa-pot}.  
Time, distance and energy are normalized to inverse of dust plasma frequency $\sqrt{2}\omega_{pd}^{-1}=\omega_{0}^{-1}$, mean inter-gain spacing $a$ and average Coulomb energy of dust particle $Q_{d}^2/(4\pi \varepsilon_{0}a)$ respectively. Therefore, all physical quantities appearing hereafter in present paper are non-dimensional. In our simulations, for a given dust particles density $\bar{n}$, the size of the system is decided by the total number of particles. For $\bar{n}= 1/\pi$, $N=2500$ and $\kappa=0.5$, we first bring our system to the desired $\Gamma$ using a Gaussian thermostat \cite{EVANS}. After this we let the system evolve under micro-canonical conditions and notice excellent agreement between temperatures obtained from both kinetic and configurational degrees of freedom (see Fig. \ref{Tc}). The absence of any noticeable drift in temperature even without a thermostat indicates good numerical stability of our time integration.

\section{Configurational Thermostat }
\label{Thermostat}
\noindent In our previous work on shear flows \citep{Aka, ashwin, ashwin_coherent}, we studied the spatio-temporal evolution of instabilities as an initial value problem. As no attempt was made to control the temperature of the liquid during the simulation, the flow evolved under adiabatic conditions and the shear heat generated due to viscosity remained in the system.  This led to a gradual increase in overall temperature eventually resulting in short lifetimes of large scale vortex structures. Thus to address macroscale vortex dynamics with longer lifetimes it is highly desirable to ``remove" this excess heat from the shear layer without altering the physics of the problem. In conventional MD simulations, thermostats are generally used to maintain the temperature of the system at a desired value in a canonical ensemble. For eg. in a typical Gaussian thermostat \cite{EVANS}, a Lagrangian multiplier is invoked for instantaneous velocities and equations of motion are augmented (in the Nose-Hoover sense) with a velocity dependent non-holonomic constraint.  While the trajectory of the system generated so forth strictly conforms to the iso-kinetic ensemble it can be shown that the observed thermodynamic behavior of the system corresponds very well to that of the canonical ensemble in thermodynamic limit. As can be expected, such velocity scaling based thermostats can work only at low shear rates and are immediately rendered useless at high shear rates where secondary flows usually develop. It is the purpose of this paper to show that a PUT can be efficiently applied to control the temperature of shear flows and below we present the details of our protocol.\\  
  
\noindent Once a macroscale flow is superimposed onto the thermalized grain bed, the instantaneous particles velocities contain information regarding the ``thermal" and the ``flow (or average)" parts. Especially at high Reynolds number regime where secondary flows usually develop, it becomes impossible to control temperature using such PBTs  which rely only on velocity scaling. Thus controlling the ``thermal" component of velocity and letting mean component evolve is impossible using thermostats which use augmented velocity equation as in a Gaussian thermostat. \textit {Is it then possible to ``thermostat" a system with $N$ particles without modifying the instantaneous velocities of particles ? The answer is yes.} As discussed earlier, a novel method of thermostatting, namely configurational thermostat has been proposed amongst others, by Braga and Travis \citep{Braga2005,BT2008}, which controls the temperature by using augmented equations of motions for the instantaneous particle positions without disturbing  the instantaneous velocity of  particles. In this PUT, instead of a non-holonomic constraint, a holonomic constraint is augmented to the equation of motion. The temperature of the system is then calculated by using configurational definition [Eq.$\:$\ref{main}] that agrees well with the kinetic temperature calculated using Eq.$\:$\ref{kinetic}.  To better understand the subtle ways in which the configurational thermostat differs from its kinetic counterpart, we describe both these schemes below.\\

\noindent The equation of motion corresponding to kinetic Nose-Hoover thermostat are: 
\begin{equation}
\dot{\boldsymbol{r}}_{i}=\frac{\boldsymbol{p}_{i}}{m_{i}}
\label{eqmk1}
\end{equation}
\begin{equation}
\dot{\boldsymbol{p}}_{i}=-\frac{\partial U}{\partial \boldsymbol{r}_{i}}-\eta \boldsymbol{v}_{i}
\label{eqmk2}
\end{equation}
\begin{equation}
\dot{\eta}=\frac{1}{Q_{\eta}}\bigg(\sum_{i=1}^{N}\frac{m_{i}v_{i}^{2}}{2}-k_{B}T_{0}\bigg)
\label{eqmk3}
\end{equation}

\noindent where $m_{i}$, $\boldsymbol{r}_{i}$, $\boldsymbol{p}_{i}$ and $T_{0}$ are mass, position, momentum of $``i"$-th particle and desired temperature respectively. Lagrange multiplier $\eta$ is a dynamical variable and Eq.$\:$\ref{eqmk1}-Eq.$\:$\ref{eqmk3} are the new augmented equations of motion. $Q_{\eta}$ is damping constant or effective mass  \citep{Noose,Nosechain}. In the same way, for $``i"$-th particle the augmented equations of motion corresponding to configurational temperature based Nose-Hoover thermostat as defined by Braga and Travis \citep{Braga2005,BT2008}  are : 
\begin{equation}
\label{Eqbt1}
\dot{\boldsymbol{r}}_{i}=\frac{\boldsymbol{p}_{i}}{m_{i}}-\mu\frac{\partial U}{\partial \boldsymbol{r}_{i}}
\end{equation}
\begin{equation}
\dot{\boldsymbol{p}}_{i}=-\frac{\partial U}{\partial \boldsymbol{r}_{i}}
\end{equation}
\begin{equation}
\label{Eqbt2}
\dot{\mu}=\frac{1}{Q_{\mu}}\bigg(\sum_{i=1}^{N}\Big(\frac{\partial U}{\partial \boldsymbol{r}_{i}}\Big)^{2}-k_{B}T_{0} \sum_{i=1}^{N}\frac{\partial^{2}U}{\partial r_{i}^{2}}\bigg)
\end{equation}

\noindent In above equations, the Lagrange multiplier $\mu$ is a dynamical variable. In  configurational thermostat,  $Q_{\mu}$ is an empirical parameter which behaves as the effective mass associated with thermostat. The value of $Q_{\mu}$ decides the strength of coupling between system and  heat-bath. We find that the value of damping constant or effective mass $Q_{\mu}$ is sensitive to the desired coupling parameter $\Gamma$ values. For a desirable $\Gamma$, we find that for low values of damping constant, the system arrives at the desired $\Gamma$ faster and vice versa. In the limit $Q_{\mu}\rightarrow\infty$ the configurational thermostat is de-coupled and formulation becomes micro-canonical ensemble. For the range of $\Gamma$ values studied here, we find that $Q_{\mu}=2\times10^{6}$ allows steady state with in time $t=100$ and hence is used throughout.\\

\begin{figure}[h!]
\includegraphics[width=3.4in,height=3.2in]{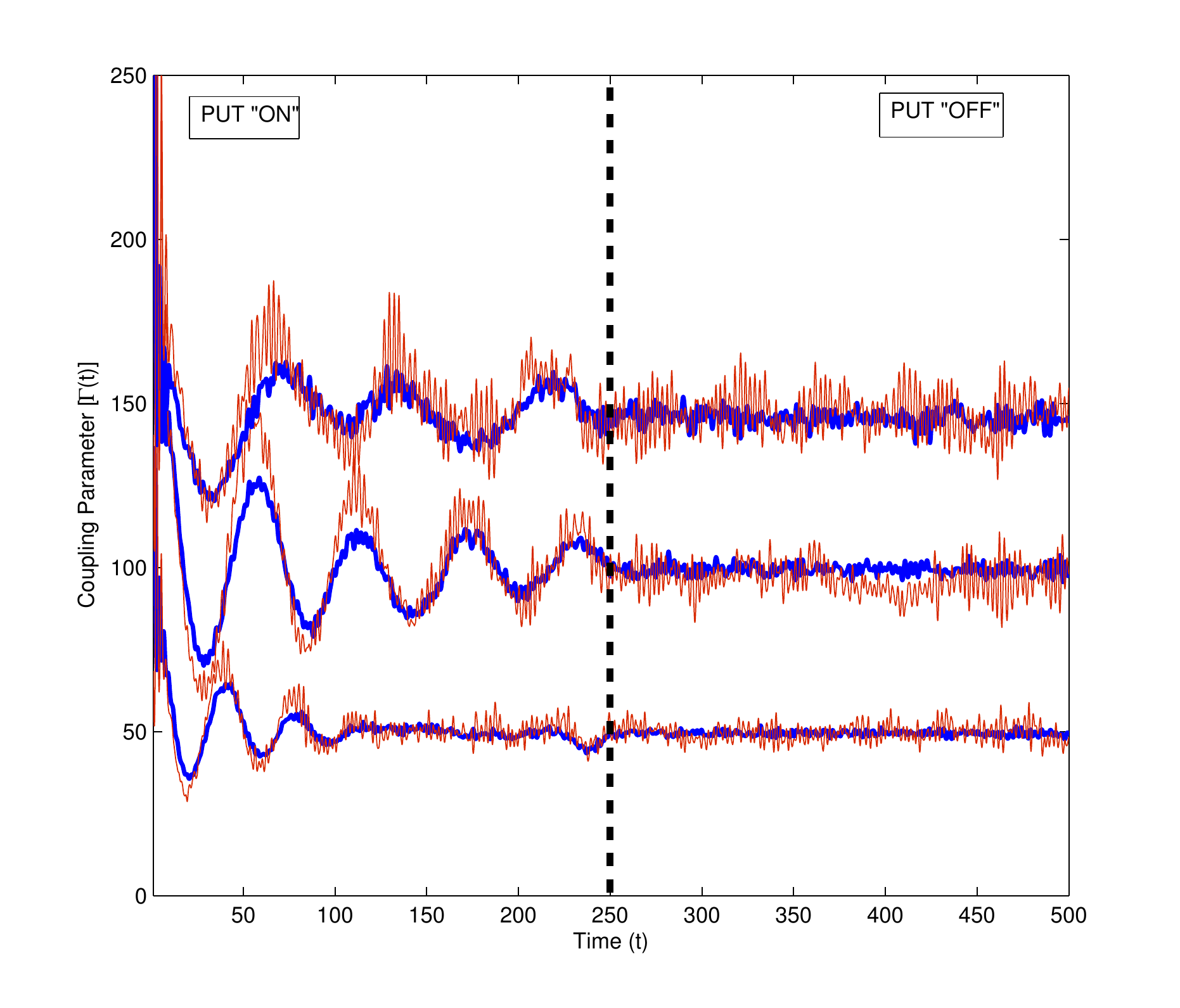}
\caption{ color online: Kinetic (blue) and configurational (red) $\Gamma$ vs time. Parameters used: $Q_{\mu}=2\times10^{6},  N=2500$ and $\kappa=0.5$.}
\label{Gammabt}
\end{figure}

\noindent Using a PUT, it is depicted in Fig.[\ref{Gammabt}] that kinetic and configurational coupling parameter $\Gamma$ follow the same behaviour in canonical and in micro-canonical run as well.
\\

\noindent In the following section, the evolution of shear flow namely, Kolmogorov flow,  in Yukawa liquids in the presence and absence of microscopic or molecular shear heat is presented.

\section{Kolmogorov Flow as an initial value problem in Yukawa liquid with and without molecular heat generation}
\label{heat}
\begin{figure}[h!]
\includegraphics[width=3.4in,height=3.2in]{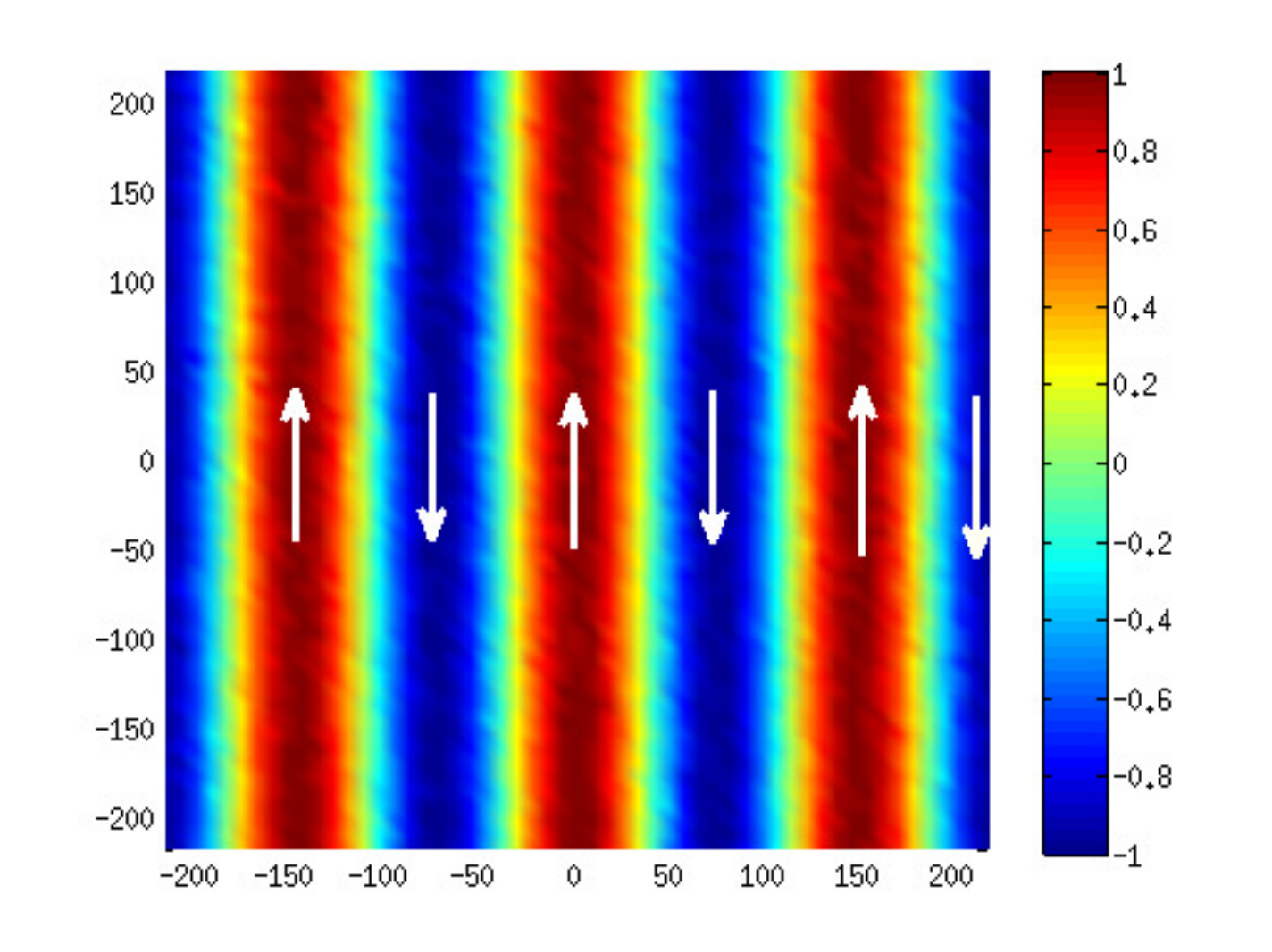}
\caption{color online: Kolmogorov velocity profile. Arrows show the direction of local flow.}
\label{KM-flow}
\end{figure}

\noindent To study the shear flow evolution and vortex dynamics from the microscopic dynamics in presence and absence of heat generation phenomena, we have studied Kolmogorov flow \citep{sinai}  as an initial value problem. This simply implies that a flow profile $\boldsymbol{U_{0}}$ is loaded only at $t=0$ and no attempt is made to control the mean flow at later times. At time $t>0$, we use a PUT to maintain the desired temperature. The loaded shear profile has the form $\boldsymbol{U_{0}}(x, y)$=$U_{0}\cos(2\pi n_{0}x/L_{x})(1+\delta\cos(2\pi my/L_{y}))\hat{y}$, where  the magnitude of equilibrium velocity $ U_{0}=1$, spatial period number $n_{0}=3$ , magnitude of perturbation $\delta=0.01$ , perturbed mode number $m=2$ [see Fig.\ref{KM-flow}]. Coupling strength at time $t=0$ is $\Gamma_{0}=\Gamma(t=0)=50$, for which calculated thermal velocity is $v_{th}=\sqrt{2/\Gamma_{0}}=0.2$, which is much smaller than the equilibrium velocity  $(U_{0}=1)$. Unlike earlier section Sec.$\:$\ref{Thermostat}, here we have considered large number of particles  $N_{d}=62500$ to  study large-scale hydrodynamic phenomena using MD simulation.  Due to the large size of the simulation box $L_{x}=L_{y}=L = 443.12$, we do not consider Ewald sums \cite{Salin}. The non-dimensional screening parameter $\kappa$ is $0.5$. It is estimated that the sound speed of the system for $\Gamma_{0}=50$ and $\kappa=0.5$ is with in the range of 2-2.5 \cite{thomas} which is larger than the equilibrium velocity $U_{0}$. Hence the shear flow is considered to be ``subsonic" in nature. \\
 
\begin{figure}
\includegraphics[width=3.4in,height=3.2in]{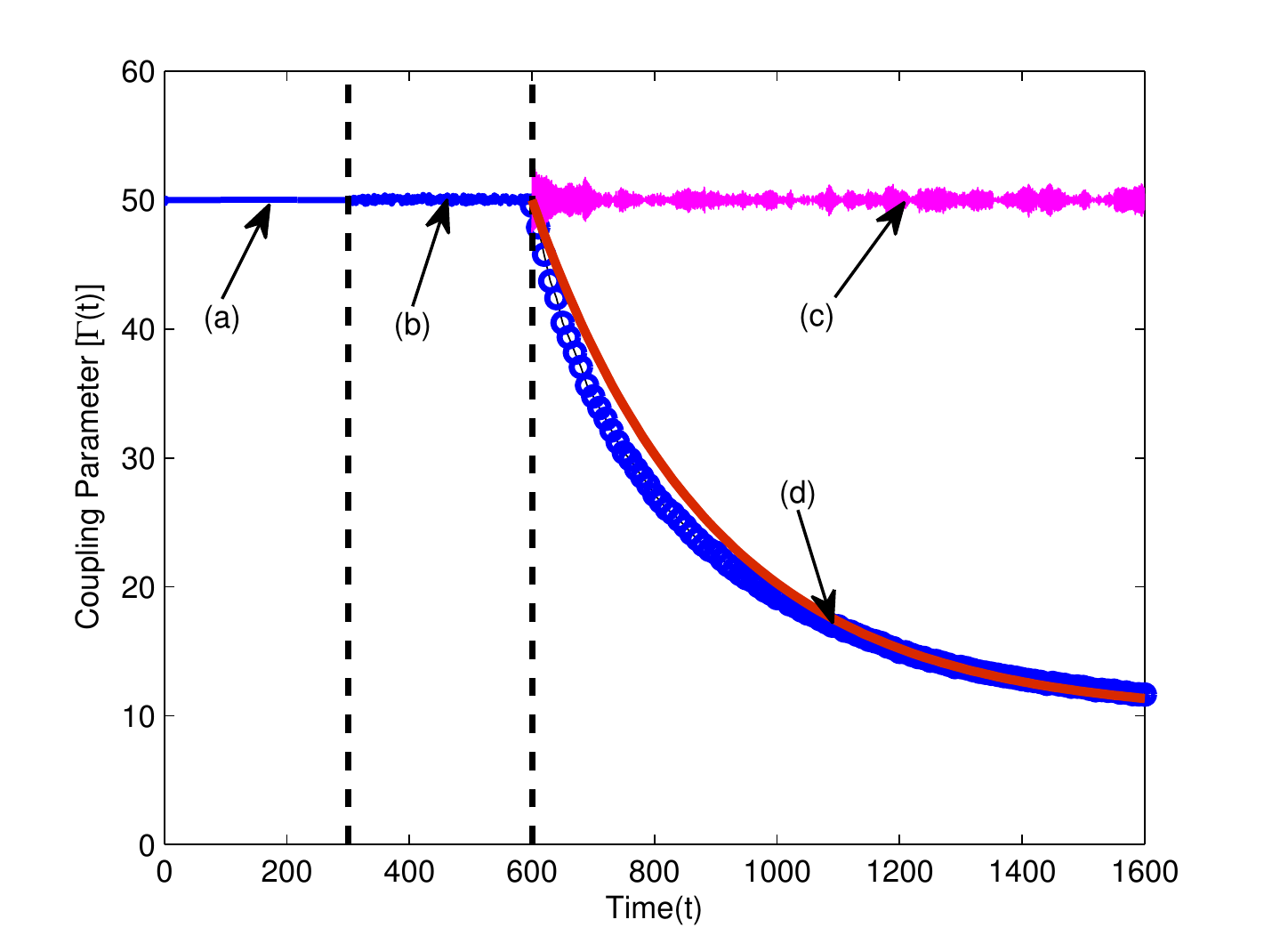}
\caption{ color online: Coupling parameter vs time. The system is evolved (a) coupled to Gaussian thermostat (b) under micro-canonical conditions (c) with PUT ``ON" and flow superposed at $t=600$ (d) with PUT ``OFF". We provide a fit (red) to show that weakening of $\Gamma$ is indeed exponential in time. Parameters used: $\Gamma_{0}=50$.}
\label{Gammashear}
\end{figure}

\begin{figure}[h!]
\includegraphics[width=3.4in,height=3.2in]{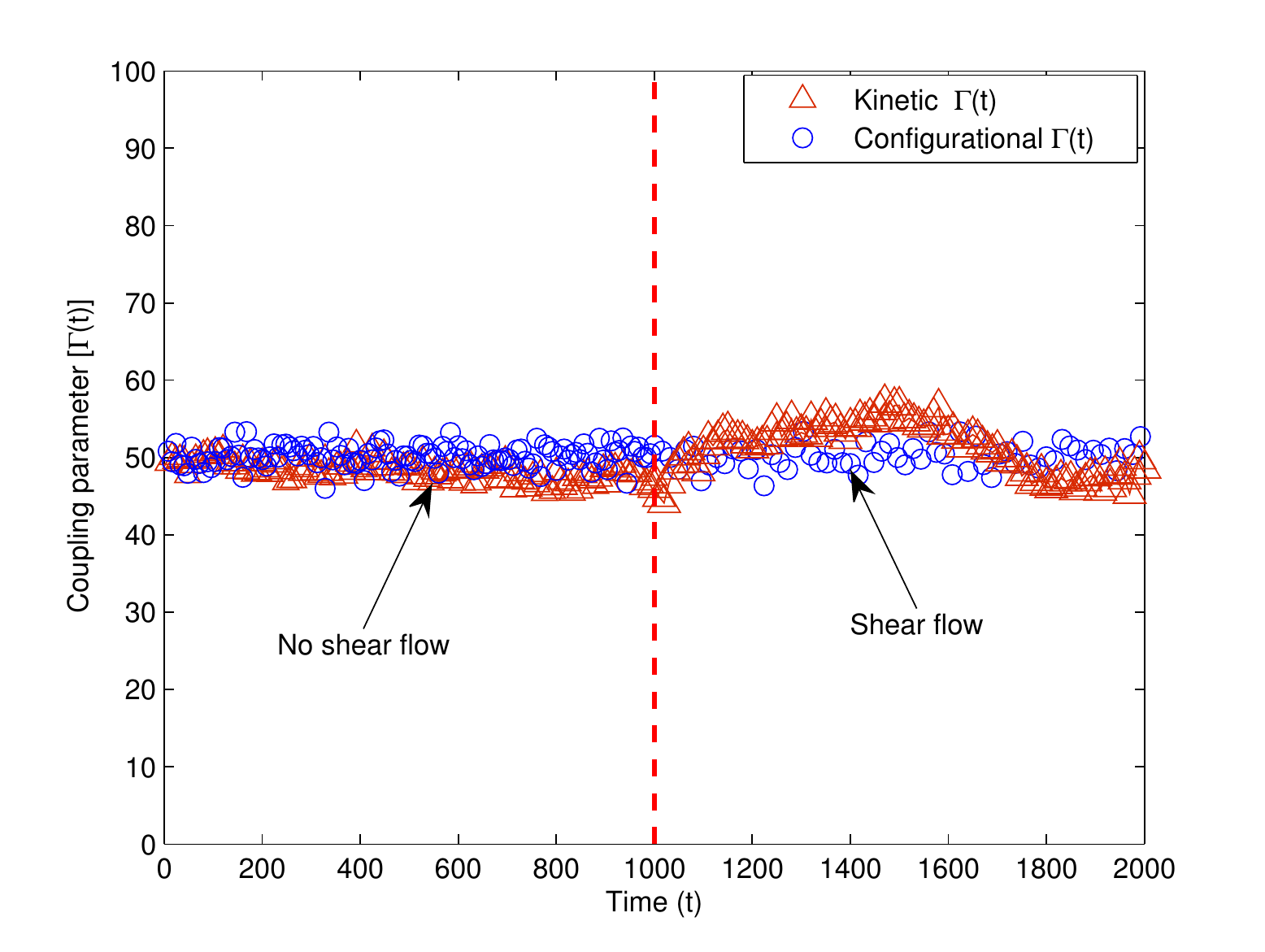}
\caption{Coupling parameter vs time plot with PUT always ``ON". Shear flow is superimposed at time $t=1000$.}
\label{GammaTc}
\end{figure}

\noindent  To see the effect of molecular shear heating, we compare our present study with the earlier work \citep{Aka}. Previously, it has been demonstrated that after superposition of the Kolmogorov flow, the coupling parameter was found to weaken under adiabatic conditions due to molecular shear heating. In the present paper, we show that this average coupling parameter when coupled to the configurational thermostat is approximately constant and close to desired $\Gamma$. In Fig.\ref{Gammashear}, we have plotted our earlier results along with the current results of average coupling parameter. In Fig.$\:$\ref{Gammashear}$(a)$  the system is thermally equilibrated up to time $t=300 $ using a conventional Gaussian thermostat. In $(b)$ we show the microcanonical run in the time interval $300<t<600$. At $t=600$ we superpose the shear flow just once and then observe the system both with PUT ``ON" $(c)$ and with PUT ``OFF" $(d)$. As clearly seen from $(d)$, the heat generated due to shear flow remains within the system and weakens the coupling strength $\Gamma$. We find this decay to be exponential in time (see fit). This is in stark contrast to the regime $(c)$ where this excess heat has been ``removed" from the system through heat-bath, which is turn facilitates $\Gamma$ to remain constant. [We repeated the study by replacing the Gaussian thermostat in $a$ with a PUT and found identical results]. We also see from Fig.\ref{GammaTc} that both the configurational and kinetic coupling parameter are close to the initial value $\Gamma_{0}=50$.

\subsubsection*{Macroscopic quantities from microscopic information (process of ``fluidization"):}
\begin{figure*}
\includegraphics[width=3.5in,height=4in]{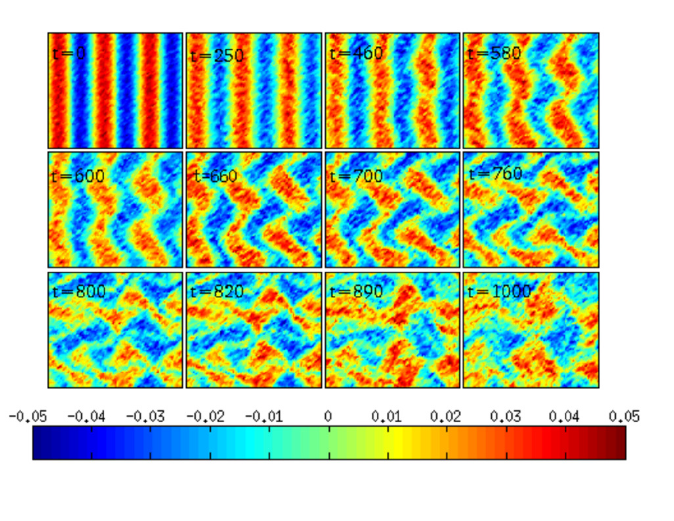}
\includegraphics[width=3.5in,height=4in]{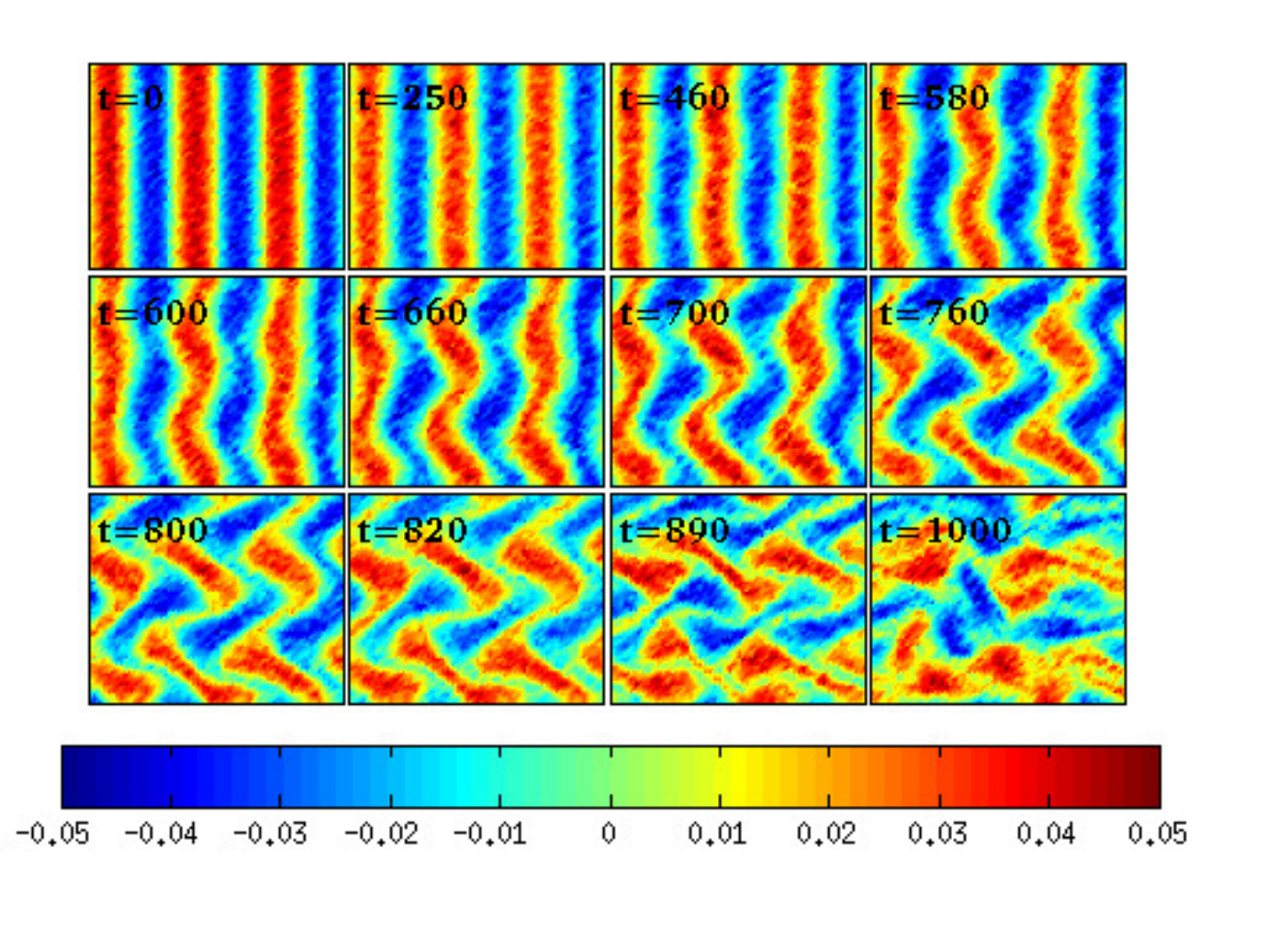}
\caption{ color online: ``Fluid" vorticity  ($\omega=\nabla\times\boldsymbol{U}$) contour plots. Color bars show the magnitude of local vorticity. Parameters used: perturbation mode $m=2$, equilibrium spatial period number  $n_{0}=3$, $\Gamma_{0}=50$, $\kappa=0.5$, initial Reynolds number $R=235.149$ and shear velocity $U_{0}$=1. Left Panel: The micro scale heating quickly destroy the vorticity structures when PUT is "OFF". Right Panel: When PUT is "ON", vortex structures sustain for longer time.}

\label{cont-vorticity}
\end{figure*}

\noindent A mesh-grid of size $55\times55$ is superimposed on the particles of the system to calculate the macroscopic or ``fluid" variables. Average local fluid velocities along $x$ and $y$ directions are calculated as  $ \bar{U}_{x}=(1/N_{b})\sum_{i=1}^{N_{b}}v_{ix}$, $ \bar{U}_{y}=(1/N_{b})\sum_{i=1}^{N_{b}}v_{iy}$, where $v_{ix}$ and $v_{iy}$ are  individual  instantaneous particle velocities along x and y direction and $N_{b}$ is the total number of particles present in an individual bin. Each bin contains approximately 20 particles [ $N_{b}=N_{d}/{(55\times55)}\simeq20$ with $N_{d}=62500$]. From average local velocities, we calculate the average local vorticity $\boldsymbol{\bar{\omega}}(x_{G}, y_{G}) = \boldsymbol{\nabla}\times \boldsymbol{\bar{U}}$ and the average local temperature $\bar{T}(x_{G},y_{G}) = (2/3)\sum_{i=1}^{N_{b}}\big((v_{ix}-\bar{U_{x}})^{2}+(v_{iy}-\bar{U_{y}})^{2}\big)/N_b$ at the Eulerian grid location $(x_{G}, y_{G})$. In Fig.[\ref{cont-vorticity}] time evolution of ``fluidized" vorticity after superposition of shear flow both in the presence (Left panel) and absence (Right panel) of PUT. We clearly see that shear heating destroy vortex structures thus resulting in their shorter lifetimes compared to the case when PUT is ``ON'' where the lifetimes of these vortices become significantly longer.\\

\begin{figure*}
\includegraphics[width=5in,height=5in]{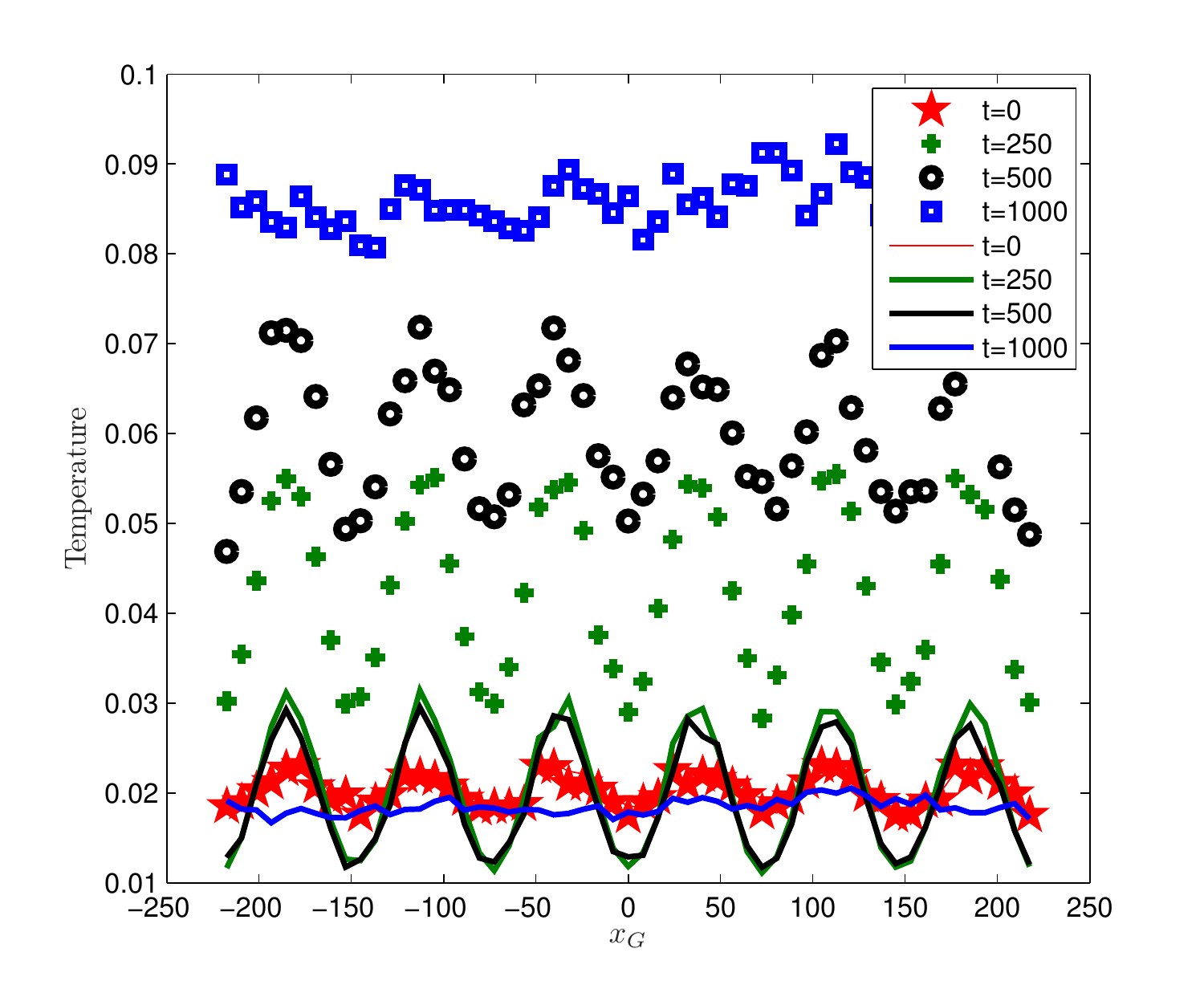}
\caption{ $y$- averaged temperature profiles ($\bar{T}(x_{G},y_{G},t)$) at various times after the shear flow is superimposed. Parameters used are  $\Gamma_{0}=50$,  $U_{0}=1$ and  $\kappa=0.5$. Symbols show the temperature profile with PUT ``OFF" and solid lines show temperature variation  with PUT ``ON".}
\label{temp}
\end{figure*}

\begin{figure*}
\includegraphics[width=3in,height=3in]{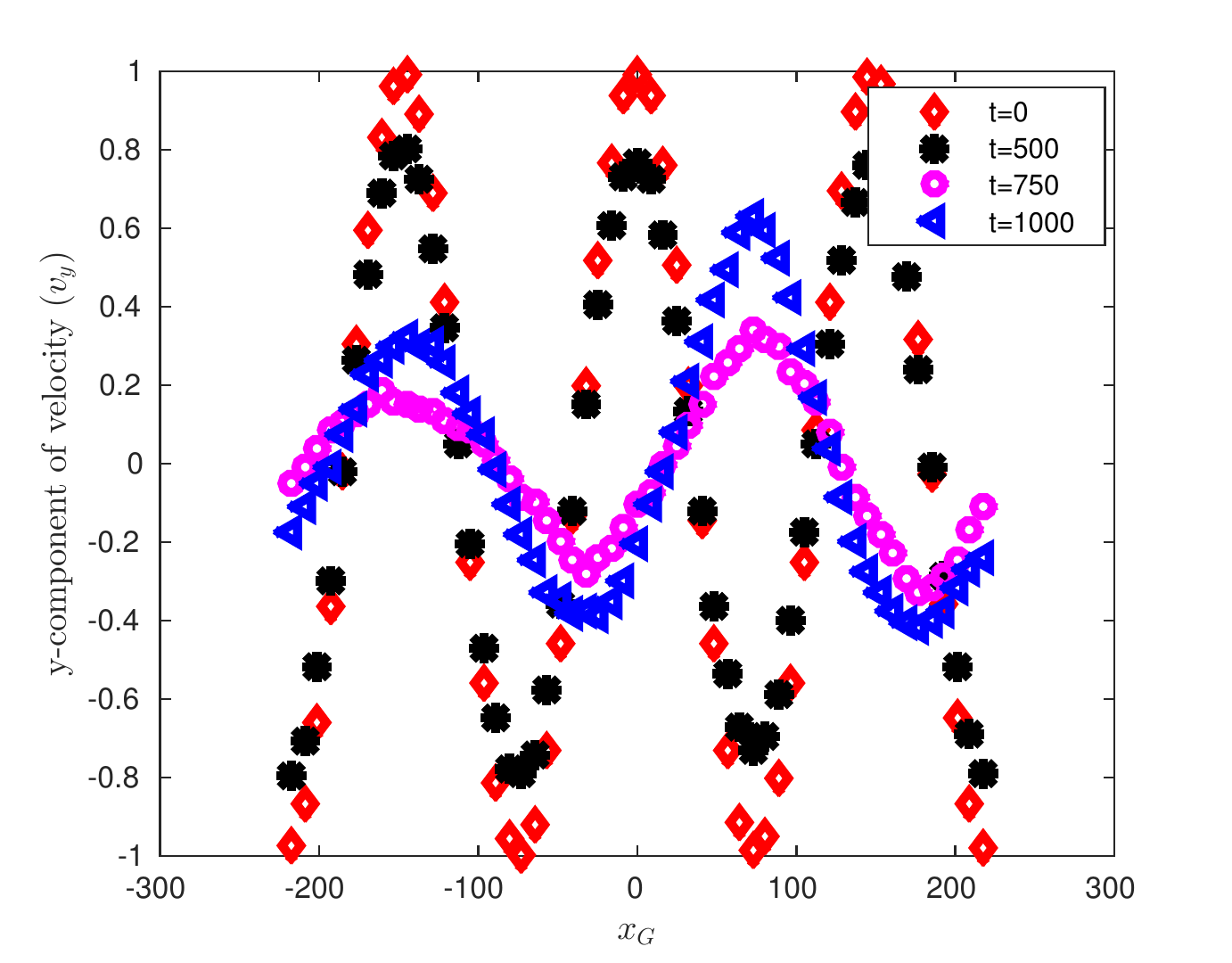}
\includegraphics[width=3.5in,height=3in]{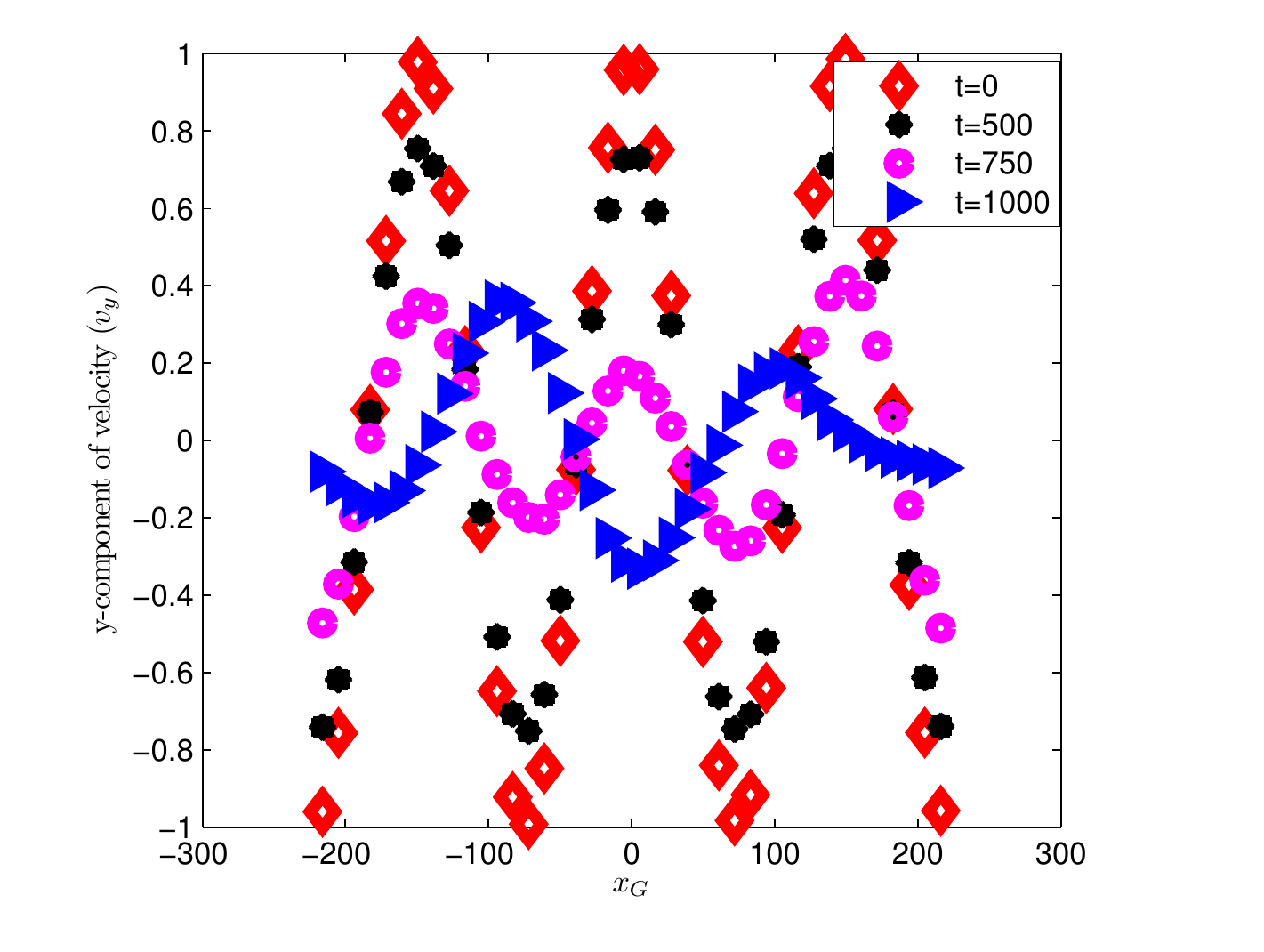}
\caption{Temporal evolution of $y$ averaged velocity $v_{y}(x_{G})$ profile for  $\Gamma_{0}=50$,  equilibrium velocity magnitude $U_{0}=1$, screening parameter $\kappa=0.5$.[Left Panel]: In the presence of heating (PUT ``OFF") [Right Panel]: In the absence of heating (PUT ``ON"). Time $t$ here are shown after superimposition of shear flow.}
\label{velocity}
\end{figure*}

\begin{figure}
\includegraphics[width=3.1in,height=3.1in]{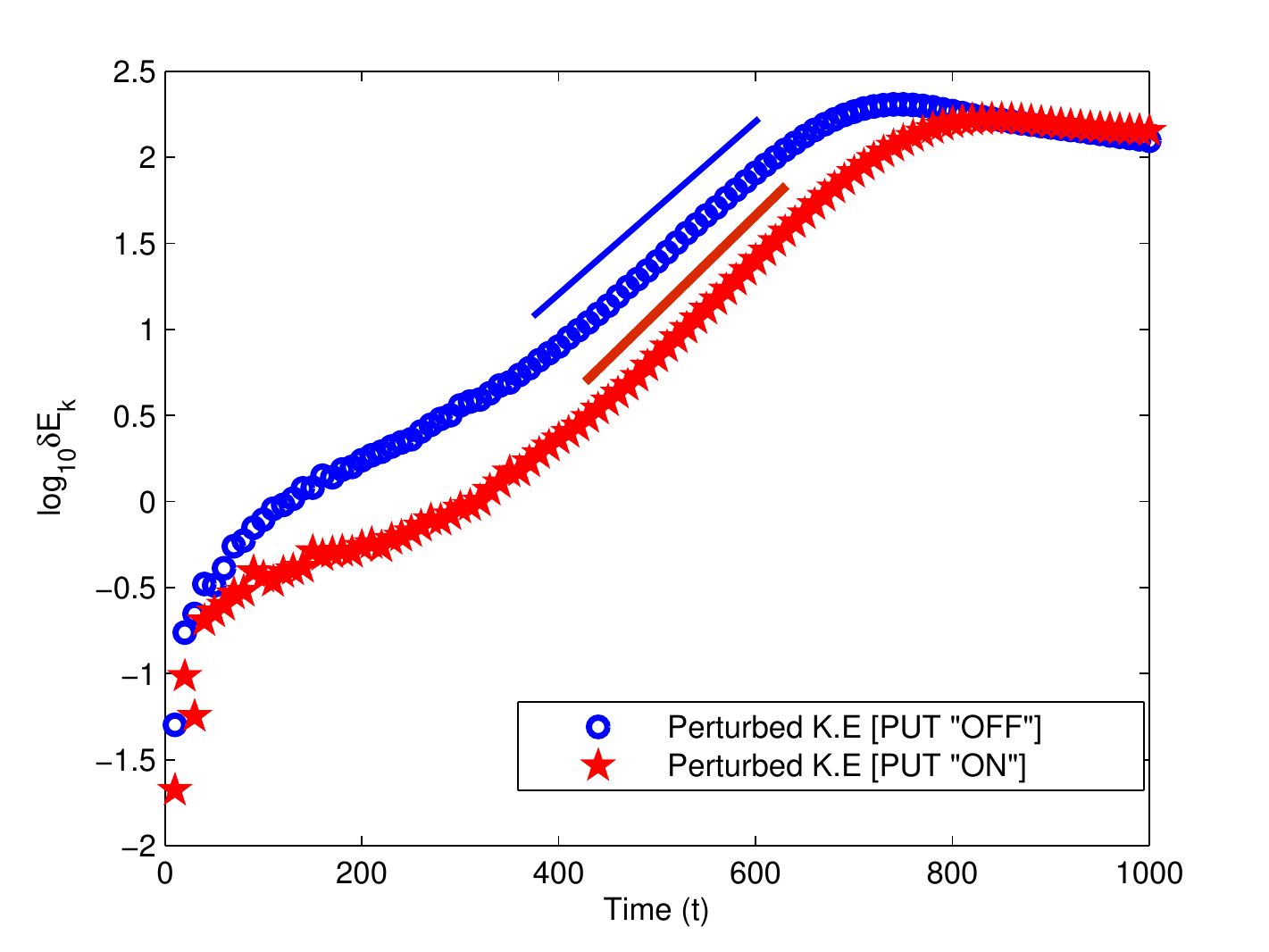}
\caption{ Perturbed kinetic energy in linear-log scale with and without PUT. Calculated growth rates from simulation are $5.5\times10^{-2}$ and $6.0\times10^{-2}$ for PUT ``OFF" and ``ON" case respectively. }
\label{Growth-gamma}
\end{figure}

\noindent  In Fig.\ref{temp}, we show our data for $\bar{T}(x_{G},y_{G})$ extracted at different times with shear flow imposed. In the absence of PUT (shown with symbols), the temperature first increases and then eventually saturates at a particular value at long times. This is in contrast to the case when PUT is present where the temperature first increases (discussed later) and then saturates at the initial temperature at long times. The value of initial temperature taken was 0.02 ($\Gamma_{0} = 50$). In Fig.[\ref{velocity}] we have plotted the $y$-component of velocity at various times in the absence (Left panel) and presence (Right panel) of PUT. We find that in both the cases the velocity profiles have not changed much between the linear and non-linear regimes of shear flow evolution. They remain qualitatively similar and differ only quantitatively. The perturbed $x$ component of kinetic energy $\delta E_{kx}$ is obtained from the expression below and we use it to calculate the growth rate shown in the Fig.[\ref{Growth-gamma}].
\begin{equation}
\biggl \lvert {\frac{\delta E_{kx}(t)}{E_{kx}(0)}} \biggr \rvert =\dfrac{\int \int [v^{2}_{x}(t)- v^{2}_{x}(0)]dxdy}{\int \int v^{2}_{x}(0)dxdy }
\label{growth_formula}
\end{equation}
 We find that the calculated growth rate $\gamma_{d}$ in the absence and presence of PUT are very close with the difference being only marginal ($<10\%$).\\

\begin{figure}[h!]
\includegraphics[width=3.5in,height=3in]{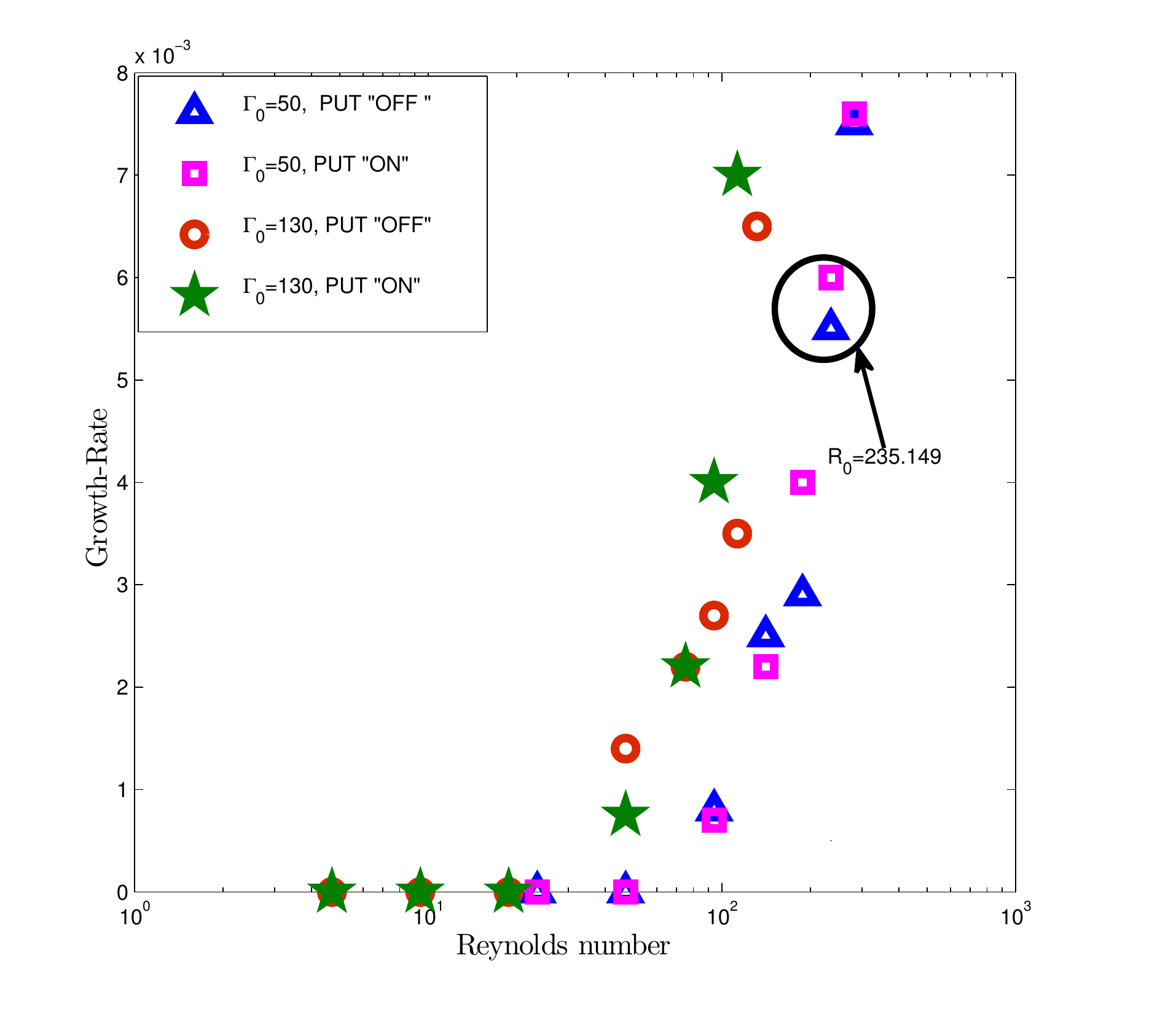}
\caption{Growth-Rate vs initial Reynolds number $\Big(R=\frac{U_{0}l\bar{n}}{\eta_{0}}\Big)$ plot showing trans-critical kind of bifurcation for screening parameter $\kappa=0.5$. The non-linear vortex structures in Fig.$\:$\ref{cont-vorticity} have been obtained for $R=235.149$ indicated here by a black circle.}
\label{bifur}
\end{figure}

\noindent In Fig.[\ref{bifur}], we have shown results of  a parametric study for maximum growth-rate of perturbed mode with initial Reynolds number $R=U_{0}l\bar{n}/\eta_{0}$, where $l$, $\eta_{0}$ are the shearing length and initial shear viscosity of the flow respectively. Here, the initial value of shear viscosity $\eta_{0}$ is calculated using the Green-Kubo formalism \citep{greenkubo, Ashwin_tau_PRL} before Kolmogorov flow superimposed. It is depicted from figure that for  given value of $\Gamma_{0}$ and $\kappa$, flow is neutrally stable below $R<R_{c}$, where $R_{c}$ is critical value of Reynolds number and for $R>R_{c}$ flow becomes unstable and eventually turbulent. Such laminar to turbulent transition in our system might be a trans-critical bifurcation \citep{drazin}. Interestingly, we find that higher values of coupling parameter $\Gamma$ decreases the critical value of Reynolds number $R_{c}$. Also the critical value of Reynolds number $R_{c}$ is found to be independent of heat generation. It is evident from Fig.[\ref{bifur}] that the growth-rate of perturbed mode ($m=2$) is not affected by molecular or microscopic heating, which shows that the suppression of heat generation does not modify the shear flow dynamics in early phase of simulation. 

\section{conclusions}
\noindent In the present paper, taking Kolmogorov flow as an initial condition, the role of molecular shear heating in strongly coupled Yukawa liquids has been investigated using configurational thermostat and results are compared  with our earlier work using micro-canonical ensemble \citep{Aka}. In both the cases, we observe that the laminar to turbulent transition of Kolmogorov shear flow crucially depends upon the critical value of Reynolds number $R_{c}$ in the presence and absence of heat generation. Parametric study of growth-rate of perturbed mode over the range of Reynolds number shows the neutrally stable and unstable nature of Yukawa fluids undergoing Kolmogorov flow for $R<R_{c}$ and $R>R_{c}$ case respectively. It is important to note that the critical value of Reynolds number in the presence and absence of heat generation is nearly same. Molecular shear heat is found to decrease the coupling strength exponentially in time and hence destroys the secondary coherent vortices. In this work, using the method of configurational thermostat, it has been demonstrated that the average or global temperature of the system can be maintained at a desirable value in spite of molecular shear heating.
This in-turn is found to help sustain the secondary coherent vortices dynamics for a longer time. In the non-linear states obtained with PUT ``ON", spatially non-uniform profiles of temperature is observed  in the regions of strong velocity shear. However, average temperature of the system is controlled by the configurational thermostat. For example when, PUT is ``ON", it is observed that  the peaks of local temperature profile at the shear flow location, are much lesser in magnitude and global average temperature of the system is maintained as compared to the case with PUT ``OFF". \\

\noindent While it has been unambiguously demonstrated here that configurational thermostats [Eq.$\:$\ref{Eqbt1}-Eq.$\:$\ref{Eqbt2}] are indeed effective in controlling the average or global temperature, it is desirable to be able to maintain the local and global temperatures at a same values. Work is under way to construct a novel set of equations of motion using an improved configurational thermostat, which will be reported in a future communication.\\

\noindent We propose that the calculation of temperature using the definition of configurational temperature may be an important and useful alternative for those laboratory experiments of strongly coupled dusty plasma in which instantaneous positions of particles are more accurately measured rather than instantaneous momenta. We strongly believe that this exercise should be undertaken using experimentally obtained instantaneous positions $\lbrace\bar{x_{i}}(t)\rbrace_{i=1,N}$ and compare the same with conventional kinetic temperature.
\label{sec.conclusion}

\begin{acknowledgments}
\noindent Authors (A.G and R.G) acknowledge  valuable discussions on configurational thermostat with Harish Charan and Rupak Mukherjee. 

\end{acknowledgments}
\nocite{*}

\providecommand{\noopsort}[1]{}\providecommand{\singleletter}[1]{#1}%

\end{document}